\documentclass[runningheads]{llncs}
\usepackage{cite}
\usepackage{amsmath,amssymb,amsfonts}
\usepackage{algorithmic}
\usepackage{graphicx}
\usepackage{balance}
\usepackage{textcomp}
\usepackage{xcolor}
\makeatletter
\newcommand*\bigcdot{\mathpalette\bigcdot@{.5}}
\newcommand*\bigcdot@[2]{\mathbin{\vcenter{\hbox{\scalebox{#2}{$\m@th#1\bullet$}}}}}
\makeatother
\begin{document}
\title{Incorporating Social-aware User Preference for Video Recommendation}
%
%
\author{Xuanji Xiao\inst{1} \and
Huaqiang Dai\inst{2} \and
Qian Dong\inst{3} \and Shuzi Niu\inst{4} \and Yuzhen Liu\inst{5} \and Pei Liu\inst{5}}
\authorrunning{Xiao et al.}
%
\institute{Shopee, Shenzhen, China\\
\email{charles.xiao@shopee.com	} \and
Xiamen University, Xiamen, China\\\email{hqdai@stu.xmu.edu.cn}\\
\and
Tsinghua University, Beijing, China\\
\email{dq22@mails.tsinghua.edu.cn}\\
\and 
Institute of Software, Chinese Academy of Sciences, Beijing, China\\
\email{shuzi@iscas.ac.cn}\\ 
\and 
Tencent Inc., Beijing, China\\
\email{\{yzhenliu,alexpliu\}@tencent.com}}
\maketitle              
\begin{abstract}
Modeling user interest accurately is crucial to recommendation systems.
Existing works capture user interest from historical behaviors. 
Due to the sparsity and noise in user behavior data, behavior based models learn incomplete and sometimes inaccurate preference patterns and easily suffer from the cold user problem.
In this work, we propose a social graph enhanced framework for behavior based models, namely \textbf{Social4Rec}. The social graph, involving multiple relation types, is extracted to find users with similar interests. It is challenging due to the trivial and sparse relations in social graph. 
To address the sparse relations issue, we first propose a Cluster-Calibrate-Merge network (CCM) to discover interest groups satisfying three properties: intrinsic self-organizing patterns through cluster layer, robustness to sparse relations through knowledge distillation of calibrator layer. 
We then use the averaged user interest representation within each group from CCM to complete each user behavior embedding and obtain relation specific interest aware embedding. 
To alleviate the trivial relation problem, relation specific interest aware embedding are aggregated among relation types through attention mechanism to obtain the interest aware social embedding for each user. It is combined with user behavior embedding to derive the matching score between the user and item. 
Both offline and online experiments on our video platform, which is one of the biggest video recommendation platforms with nearly one billion users over the world, demonstrate the superiority of our method, especially for cold users. The codes are available at https://github.com/xuanjixiao/onerec.

\end{abstract}

\section{Introduction}
Recommendation systems play a vital role in contemporary content platforms, such as video streaming and news websites, by aiming to provide users with relevant and personalized content. The main challenge of recommendation systems is to identify user interest patterns for building effective recommendation models. Existing approaches mainly depend on capturing user interests from historical behaviors, such as clicks, views, and other interactions. Some examples of these approaches are Wide\&Deep \cite{cheng2016wide}, DIN \cite{zhou2018deep}, NeighborDNN \cite{xiao2023neighbor} and the classic video recommendation model YouTubeDNN \cite{covington2016deep}. 

However, user historical behaviors are frequently \textbf{sparse} and \textbf{noisy}. Users typically engage with a small subset of items they are truly interested in, leading to sparse behavior data. The sparsity of behavior data hampers the accurate capture of user interests and preferences~\cite{xiao2020lt4rec,ouyang2023click,li2023stan,gope2017survey,sethi2021cold,bobadilla2012collaborative,lu2020meta,zhang2019contextual,li2017user}. 
The noise in historical behaviors arises from multiple biases that impact user actions. For example, users often click on items at the top of the recommendation list, resulting in a bias towards popular or highly ranked items. Since user interests are indirectly manifested in their behaviors, these behaviors alone do not sufficiently capture the complexity of user interests. This limitation arises from the homogeneity of information from a single source, resulting in the failure to capture the diverse and nuanced aspects of user preferences. 
Furthermore, behavior-based models frequently face challenges when serving \textbf{cold} \textbf{users}.
Cold users refer to users who have recently joined the platform or have shown limited engagement with the content. 
Due to the limited or absence of behavioral data, traditional behavior-based models fail to effectively comprehend the preferences of cold users and offer accurate recommendations. However, it is crucial to address the needs and preferences of cold users for the continuous growth and development of content recommendation platforms.

Consequently, recommendation models that solely rely on historical behavior data often learn incomplete and inaccurate user interest patterns. These models are likely to face difficulties in delivering satisfactory recommendations, particularly for cold users. 
An approach that considers the 
A recommendation method that can incorporate multiple heterogeneous information sources is necessary, to alleviate the sparsity and noise limitation in user behavior data and tackle the challenges presented by cold user scenarios.
\begin{figure*}
		\centering
		\includegraphics[width=\textwidth]{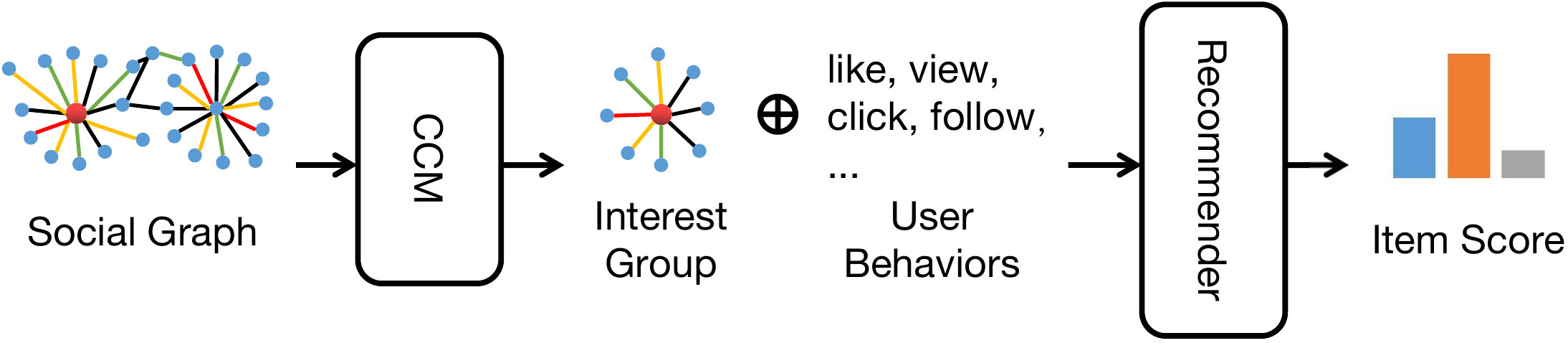}
		\caption{The workflow of Social4Rec.}
		\label{fig:workflow}
\end{figure*}
To address the limitations of behavior based models mentioned above, we seek to utilize the social graph to help complete interest patterns from behavior based models. Social graphs on modern content platforms, such as Tencent, TikTok, and YouTube, are comprehensive and easily accessible. An social graph example in Fig.~\ref{fig:workflow}, is composed of different types of relations between users, such as friendship and following the same superstars, with different colors. The red and blue nodes represent target user and its neighbors respectively. The distance between nodes indicates the degree of interest consistency between users (i.e., the closer the more consistent of interests). Intuitively, similar interest preferences may exist among users when they are actual friends on our platform, follow the same superstars, subscribe to the same movie topics, follow the same uploaders, etc. Therefore, the social graph could be beneficial for RS. Despite the rich information contained in the social graph, it is difficult to take advantage of it directly for behavior based models due to the following two challenges:
 


\begin{itemize}
\item \textbf{Trivial Relation Challenge.} 
The social graph contains trivial relations, i.e., some neighbors are distant from the target user in Fig.~\ref{fig:workflow}, which hardly implies the interest characteristics of the target user. The inappropriate incorporation of the social graph could even jeopardize the RS performance. How to utilize the social graph with trivial relations in RS is remain a challenge in real RS.

\item \textbf{Sparse Relation Challenge.} 
Despite the relations are abundant in the social graph, the types of edge (i.e. the types of relation between users) still suffer from the sparse issue. Most users only have one type of neighbor. How to learn interest preferences effectively from social data with sparse relation types is still a challenge.
\end{itemize}

To tackle the above challenges, we propose a novel social graph enhanced framework for behavior based recommendation, namely \textbf{Social4Rec}. The whole framework is composed of two modules in Fig.~\ref{fig:workflow}. One is to discover user interest groups from social graphs in face of these two challenges by the proposed Cluster-Calibrate-Merge network, namely CCM. The other is to refine user embedding based on learned interest groups from CCM, and matching it with item embedding, namely social enhanced recommendation. 

Specifically, Cluster-Calibrate-Merge network is utilized to find user interest groups, which includes a cluster layer, calibrator layer and merge layer sequentially. The cluster layer is mainly based on self-organizing network~\cite{kohonen2007kohonen,carpenter1988art,seiffert2001self}, which discovers the intrinsic clustering patterns by competitively learning the network parameters. Taking the output of the cluster layer for initialization, a knowledge distillation technique is adopted to find a more robust and similar group assignment to the initialization through the calibrator layer in face of Sparse Relation Challenge. The merge layer is to find an interest consistent group assignment without too small groups by k-means. 

Based on interest group assignments obtained from CCM, we further refine user embedding from behavior based models in terms of group and relation type. For the group level of each relation type, the user embedding is averaged to complete each user embedding in this group and obtain relation specific interest aware social embedding for each user. For the relation level, all the relation specific interest aware social embedding is aggregated through attention mechanism to tackle the Trivial Relation Challenge. Finally, the derived interest aware social representation and user behavior embedding are concatenated and fed into a vanilla recommender to calculate the matching score for each candidate item. Both offline and online experiments on our video platform, which is one of the biggest video recommendation platforms with nearly one billion users over the world, demonstrate the superiority of our work, especially for cold users. 

Overall, our contributions can be summarized as follows:
\begin{itemize}
\item We propose a novel social graph enhanced paradigm to tackle limitations of behavior based models. As far as we know, it is the first attempt to introduce the social graph into large-scale online RS.
\item In face of Sparse Relation Challenge, we design a Cluster-Calibrate-Merge network to discover intrinsic, robust and interest consistent groups. Base on learned group information, user behavior embedding is further refined through attention mechanism in face of Trivial Relation Challenge.
\item Experimental results on both offline and online demonstrate the superiority of Social4Rec, especially for cold users, over the best baseline in our video recommendation platform. 
\end{itemize}

\section{Related Work}
\subsection{Behavior-based Recommendation}
Behavior-based recommendation, which is an emerging topic in RS, has attracted a wealth of researchers from both academia and industry.
YouTubeDNN~\cite{covington2016deep} is the most classic video recommendation model which has been deployed in many industrial video recommendation platforms. 
deepFM~\cite{guo2017deepfm} and xdeepFM \cite{lian2018xdeepfm} jointly learn explicit and implicit feature interactions effectively without feature engineering.
DIEN~\cite{zhou2019deep} and UBR4CTR \cite{qin2020user} model the recommendation task from the perspectives of user behavior evolving and similar behaviors retrieval respectively.

\subsection{Knowledge-enhanced Techniques}
Researchers have devoted substantial efforts to knowledge-enhanced techniques and benchmarks due to the superiority of diversified knowledge from different domain~\cite{sun2019ernie,su2021cokebert,dong2022incorporating,xie2023t2ranking}. 
Both latent knowledge~\cite{dong2021latent,dong2021legal,li2023sailer,dong20233,dong2022disentangled} and explicit knowledge~\cite{su2021cokebert,dong2022incorporating,cheng2023layout} are explored by researchers.
Recently, considerable knowledge-enhanced RS have been proposed for accurate interest modeling~\cite{xia2021graph, huang2021knowledge, xia2021knowledge, xia2022multi,yang2022knowledge}.
However, existing knowledge-enhanced RS mainly utilize knowledge graph of user-item, which is limited in the homogenized data. In this work, we propose a novel paradigm to enhance the RS with the interest knowledge distilled from social graph, i.e., user-user graph. 

\section{Preliminary}
\subsection{Self-organizing Neural Network}
\label{sec:sonnintro}
Through an unsupervised competitive learning mechanism, Self-organizing Neural Network (SoNN) could discover the intrinsic patterns from data by self-adjusting the network parameters~\cite{kohonen2007kohonen,carpenter1988art,seiffert2001self}.
Formally, given a user embedding, SoNN assigns him/her to an interest group $j$ as follow
\begin{equation}
	\underset{j}{arg \ min} \left\|{\mathbf{W}}_{j} -f\left({X_u}\right)\right\| \quad\left( j=1,2, \cdots, m\right),
\label{eq:sonnAssignment}
\end{equation}
where $X_u\in\mathbb{R}^{d}$ is the embedding of user $u$ elaborated in the following section.
$f(\cdot)$ are full-connected layers.
The Eq.~\ref{eq:sonnAssignment} measures the interest consistency between a user and an interest group by the Euclidean distance between the user $X_u$ and the interest group $\mathbf{W}_j$, where $\mathbf{W}\in\mathbb{R}^{m\times d}$ and $m$ is the number of interest group.
After a user $u$ is assigned to an interest group $j$, the embedding of the interest group $\mathbf{W}$ is updated as follows $\mathbf{W} = \mathbf{W}+d\mathbf{W}$. Despite the group embedding $\mathbf{W}$ are initialized randomly, the final embedding still reflects the intrinsic characteristics of the interest group properly after several iterations in SoNN.

Moreover, the user embedding $X_u$ is updated through back propagation based on minimizing the following loss function
\begin{equation}
\mathcal{L}_u =\sum_{u\in \mathbb{U}} \left\|\mathbf{W}_j-f\left(X_u\right)\right\|^2,
    \label{eq:sonnUserLoss}
\end{equation}
where $\mathbb{U}$ is the user set in system.
After multiple iterations, each interest group and user obtains a stable embedding, which can be employed for unsupervised user clustering accurately.
\subsection{Recommendation}
Given a user embedding $F_u$ obtained from its historical behaviors, a qualified recommender calculates the relevant score between the target user $u$ and candidate item set $\mathbb{T}$ in system, and returns top-relevant items that may be of interest to the user $u$. The embedding $F_t$ of item $t \in \mathbb{T}$ is obtained from its features, such as the category item $t$ belongs to, the number of item $t$ has been liked, etc.
The relevant score $\bar{y}_{u,t}$ between user $u$ and item $t$ is calculated by recommender as follow
\begin{equation}
\bar{y}_{u,t}=h(F_u)\bigcdot g(F_t),
    \label{eq:recommender}
\end{equation}
where $h(\cdot)$ and $g(\cdot)$ are full-connected layers or other arbitrary neural network modules, and $\bigcdot$ means the inner product operation. This is a commonly used network backbone of commercial recommendation systems.
\section{Methodology}
\begin{figure*}
		\centering
		\includegraphics[width=\textwidth]{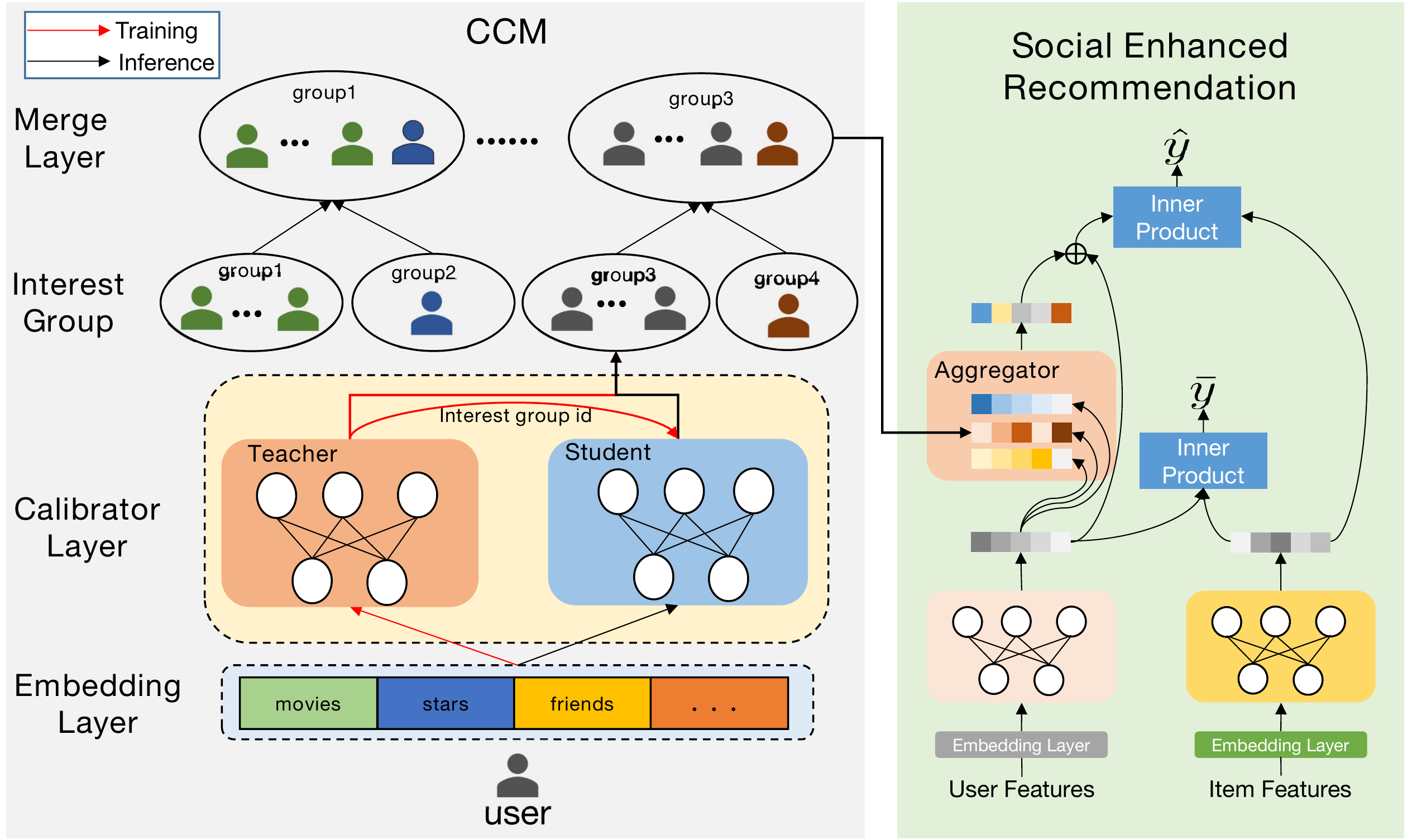}
		\caption{The architecture of Social4Rec.}
		\label{fig:architecture}
\end{figure*}
In this section, we present our method in details, which introduces the \textbf{social} graph to enhance the user interest modeling \textbf{for} better video \textbf{rec}ommendation, namely \textbf{Social4Rec}. 
Social4Rec is mainly composed of two modules. The Cluster-Calibrator-Merge module is proposed to find intrinsic, robust and self-consistent social sub-graph (i.e., interest group) structures based on self-organizing, knowledge distillation and k-means techniques. The other is social graph enhanced recommender module, which refines the user behavior embedding based on group interest characteristic distilled from Cluster-Calibrator-Merge module.


\subsection{Cluster-Calibrator-Merge Module}
\label{sec:sonn}
Birds of a feather flock together, and thus users with similar interests should be divided into one group. 
The potential interest of a user may be embedded in other users in his/her interest group.
However, most neighbors of a user in the global social graph are trivial, as for example the interest preferences of users who subscribe to the same movie topics still vary greatly.  
Therefore, the global social graph should be carefully refined into an interest group, which is robust and self-consistent for the downstream user representation learning. 
In this section, we propose a Cluster-Calibrate-Merge module, which is utilized to unsupervised cluster the users in social graph into several interest groups.
Specifically, the users in social graph are first encoded through the embedding layer. Then, each user is assigned a virtual interest group through a calibrator. To alleviate the issue that users are sparse in some interest groups, we employ K-means clustering scheme to merge interest groups containing few users into adjacent interest group.

\textbf{Embedding Layer.} As illustrated in Fig.~\ref{fig:workflow}, we define that users have a $movie$ relation if they subscribe to the same movie topic, a $star$ relation if they follow the same star, or a $friend$ relation if they are actual friends on the platform, etc. More formally, given a user with its information used for social graph construction,
we can formulate the embedding layer as follow
\begin{equation}
    X_{u} = 
    \frac{1}{n_1^{u}}\sum_{i}^{n_1^{u}}e_i^1 \oplus \frac{1}{n_2^{u}}\sum_{i}^{n_2^{u}}e_i^2 \oplus \cdots \oplus \frac{1}{n_L^{u}}\sum_{i}^{n_L^{u}}e_i^L,
    \label{eq:socialemblayer}
\end{equation}
where $L$ represents the number of relation types (e.g., $movie$, $star$, $friend$, etc.).
$n_l^u$ represents the number of specific entities in relation type $l$ of user $u$.
$e_i^l$ represents the embedding of the $i$-th specific entity in relation type $l$ (e.g., Michael Jackson and Jay Chou are two specific entities of relation type $star$). 
We use the concatenation operation $\oplus$ in embedding layer due to the heterogeneity of the specific entities between different relation types. Notably, the number of relation types is less than $L$ for most users. For the absent relation types, we use zero to pad corresponding positions in $X_u$.

\textbf{Cluster Layer.} According to Self-organizing network, the group assignment $f(\cdot)$ is updated by optimizing Eq.~\ref{eq:sonnAssignment} given user embedding and the user embedding $X_u$ is updated by optimizing Eq.~\ref{eq:sonnUserLoss}. Two steps are alternatively updated for each iteration. After multiple iterations, the process convergence and each interest group and user obtains a stable embedding. The learned group assignment $f(\cdot)$ serves as a guide for next layer.

\textbf{Calibrator Layer.} To mitigate the discrepancy introduced by the \textbf{Sparse Relation Challenge} in real social graph, we design a novel calibrator inspired by Masked Language Model (MLM)~\cite{devlin2018bert} in the community of natural language model.
Meanwhile, with the light of distillation techniques~\cite{yang2022cross, huangtwo}, we utilize the well-trained $f(\cdot)$ from Section \ref{sec:sonnintro} which is trained on the subset $\overline{\mathbb{U}}$ of users with all relation types as the teacher model, and then train a student model $k(\cdot)$ which is more robust on the social graph with sparse relations. Particularly, the input embedding $X_u$ of teacher model is polluted as $\check{X}_u$ for the student model, where a small fraction of its relations in Eq.~\ref{eq:socialemblayer} are randomly replaced by zero. 
The goal of calibrator is to mimic the predicted distribution between teach $f(\cdot)$ and student $k(\cdot)$, where the loss can be
measured by KL divergence as
\begin{equation}
\mathcal{L}_k =\sum_{u\in \overline{\mathbb{U}}} f(X_u)\cdot log\frac{k(\check{X}_u)}{f(X_u)}.
    \label{eq:klloss}
\end{equation}

\textbf{Merge Layer.}
There is a trade-off between the number of interest group and the interest consistency of users within an interest group. To ensure that model can model accurate interest characteristics from social graph, we prioritize maintaining interest consistency within an interest group. Consequently, some interest groups have few users which also needs to be avoided. To address this issue, we employ the k-means clustering scheme to merge interest groups that contain few users with the nearest interest group. 

Through the above modules, the CCM is empowered with the capability of modeling users with sparse relations in social graph and dividing users into several interest groups accurately.
\subsection{Social Enhanced Recommendation Module}
\label{sec:SER}
After obtaining the user's corresponding interest group, we first integrate the neighbors of target user $u$ within its interest group by relation types, which can be formulated as
\begin{equation}
H_u^l=avg(\sum_{\mu \in N_u^l} X_{\mu})+ X_u,
    \label{eq:aggUsers}
\end{equation}
where $N_u^l$ is the neighbors with the relation type $l$ to user $u$.
Intuitively, different users have different interest consistency with neighbors of different relation types. To tackle the \textbf{Trivial Relation Challenge} in social graph, we employ the attention mechanism to finalize the overall interest-aware social representation of user $u$ as follow
\begin{equation}
H_u=\sum_{l\in L}\alpha_{u,l}\times H_u^l,
    \label{eq:aggFinal}
\end{equation}
where $\alpha_{u,l}$ is the attention score calculated by
\begin{equation}
\alpha_{u,l}=\frac{exp(\sigma(\beta_{u,l}))}{\sum_{\iota \in L}exp(\sigma(\beta_{u,\iota}))},
    \label{eq:attnScore}
\end{equation}
where $\sigma$ is an activation function and the logits $\beta_{u,\iota}$ is computed as
\begin{equation}
\beta_{u,\iota}=q(F_u\oplus H_u^{\iota}).
    \label{eq:attnLogits}
\end{equation}

In Eq.~\ref{eq:attnLogits}, $q(\cdot)$ and $F(u)$ are full-connected layers and behavior-based embedding of user $u$ respectively. Guided by the behavioral characteristics which are more accurate and easier to interest learning, a reliable social representation $H_u$ could be obtained for the supplement of interest characteristics. With the interest-aware social representation $H_u$, the Eq.~\ref{eq:recommender} could be enhanced by the social graph $\mathcal{G}$ as follow
\begin{equation}
\hat{y}_{u,t,\mathcal{G}}=h(F_u \oplus H_u)\bigcdot g(F_t),
    \label{eq:socialEnhancedrecommender}
\end{equation}
where $\hat{y}_{u,t,\mathcal{G}}$ represents the relevant score that is aware of the interest characteristics extracted from $\mathcal{G}$.

\section{Experiments}
Our experiments are guided by the following research questions:
\begin{itemize}
\item \textbf{RQ1:} Whether the existing RS perform worse on cold users?
\item \textbf{RQ2:} Does Social4Rec enhances the performance of existing RS significantly?
\item \textbf{RQ3:} Does the deployment of Social4Rec contribute to the growth of our video platform's user base?
\end{itemize}
\subsection{Experimental Setup}
\textbf{\emph{Dataset}}. 
Existing datasets~\cite{ben2015recsys,harper2015movielens,meyffret2012red,ni2019justifying,wan2016modeling} used for recommendations contain only user behavior data and no user corresponding social data, which is readily available in a mature content platform, such as Tencent, TikTok, and YouTube.
Therefore, we conduct experiments in our video platform, which is a real-world video recommendation platform with nearly one $\emph{billion}$ users over the world.
The detailed statistics of social graph and video recommendation datasets used in this work are reported in Table~\ref{tab:summaryOfSG} and Table~\ref{tab:summaryOfVideo} respectively. For the \emph{Social Graph}, we construct it from four relation types including $star$, $movie$, $friend$ and $video$ $uploader$. For the \emph{FULL} dataset, we collect 15 consecutive days' online traffic log in our video recommendation platform, with the first 14 days for training and the last day for offline testing. For the \emph{COLD} dataset, we process the \emph{FULL} dataset by removing users with more than 30 historical behaviors in traffic log, as we find that users with less than 30 behaviors have a significant drop in impression of video items. In this dataset, 80\% of users have less than two video clicks, and thus it is hard to mine interest preference from historical behaviors.
\begin{table}[!t]
\small
\centering
\caption{Statistics of social graph.}
\begin{tabular}{cccc}
\hline
stars  & movies  & friends  & uploaders  \\
\hline
6M   &  63M  &  65M  & 25M  \\
\hline
\end{tabular}
\label{tab:summaryOfSG}

\end{table}

\begin{table*}[!t]
\small
\centering
     \caption{Statistics of the recommendation datasets.}
      \begin{tabular}{lcccccc} 
    \hline
Dataset  & user  & video  & impression & click &\#impression/user &\#click/user \\
\hline
\emph{FULL}      &  68M   &  440M  &  340M  & 136M &5.0&2.0 \\
\emph{COLD}      &  2M   &  4M  &  3M  & 1M &1.5&0.5 \\
\hline
\end{tabular}
\label{tab:summaryOfVideo}
\end{table*}
 


\noindent \textbf{\emph{Baselines.}} 
The vanilla behavior-based RS achieved the best performance in our video platform is employed as a primary baseline in both offline and online experiments, which utilizes YouTubeDNN~\cite{covington2016deep} as architecture backbone. 
The user and item features are mined by ourselves. 
It is worth noting that in real video recommendation systems, such as YouTube, TikTok, and Tencent, YouTubeDNN~\cite{covington2016deep} is commonly selected as the backbone of the recommendation model due to its simplicity and effectiveness. These video platforms typically maintain their own user and item features to ensure adaptation to their respective platforms.
To be consistent with the baseline, YouTubeDNN~\cite{covington2016deep} is also utilized as the backbone of Social4Rec, i.e., $h(\cdot)$ and $g(\cdot)$ in Eq.~\ref{eq:socialEnhancedrecommender}.
Social4Rec$^{\dagger}$ and Social4Rec$^{\ddagger}$ represent the versions of Social4Rec without calibrator and merge layer respectively.
Social4Rec$^{-}$ directly aggregates $H_u$ by averaging without using the attention aggregator. 
The Eq.~\ref{eq:aggUsers} of Social4Rec$^{-}$ could be redefined as
\begin{equation}
H_u=avg(\sum_{l \in L}H_u^l).
    \label{eq:redefinedAggFinal}
\end{equation}

\noindent \textbf{Parameter setting.}
In our experiments, we have chosen to employ the Adam optimizer~\cite{kingma2014adam} with a fixed learning rate of 0.001. To improve the stability and convergence speed of our neural networks, we have applied batch normalization in each layer. Batch normalization normalizes the activations of each layer by subtracting the batch mean and dividing by the batch standard deviation. This technique helps to address the internal covariate shift problem and allows for smoother optimization, enabling faster training and better generalization performance. We have chosen the leaky ReLU activation function for all layers, except for the last layer, which utilizes the sigmoid activation function. The leaky ReLU activation function is an extension of the traditional rectified linear unit (ReLU) function, introducing a small slope for negative input values. This helps to mitigate the issue of "dying ReLUs" by allowing a small gradient to flow through when the neuron is inactive. The sigmoid activation function in the last layer is commonly used for binary classification tasks, as it squashes the output into the range of zero to one, representing the probability of the positive class. By adopting these settings as shared between all our baseline models, we aim to establish a consistent experimental setup, allowing us to focus on the impact of other variations or enhancements on the performance of our models.


\noindent \textbf{\emph{Metrics.}} 
For the performance metrics, we resort to the widely used ${AUC}$ (Area Under Curve) as the offline metric. AUC measures the quality of the ranking produced by the recommendation system, indicating the probability that a randomly chosen positive example (e.g., a clicked video) is ranked higher than a randomly chosen negative example (e.g., a non-clicked video). A higher AUC value implies better discrimination between positive and negative examples.

In addition to the offline metric, we have incorporated online metrics to capture the real-world performance of our method. One of the key online metrics we consider is the $CTR$ (Click-Through-Rate). CTR measures the ratio of the number of clicks on recommended videos to the number of impressions (the times the recommendations are shown). A higher CTR signifies that our recommendations are attracting more user attention and engagement. To gain further insights into the user experience, we also track the $click$ $number$ and $view$ $time$. The click number refers to the total number of clicks made by users on the recommended videos, providing an indication of user interest and engagement. The view time represents the total time spent by users watching the recommended videos. Analyzing these metrics helps us understand whether cold users, who may have limited interaction history with the platform, are benefiting from our recommendations and finding content that matches their preferences.

To ensure fair and consistent comparisons across our experiments, we have set the same configuration of evaluation throughout. By evaluating both offline and online metrics, we can gain a comprehensive understanding of the quality of our method, as well as the user engagement and satisfaction with our video recommendation platform.

\subsection{Experimental Results}
The offline experimental results on $\emph{FULL}$ and $\emph{COLD}$ datasets are illustrated in Table \ref{tab:offline}. We compare the performance of the Social4Rec model against the Vanilla RS (behavior-based RS) on both datasets. The results show that Social4Rec outperforms Vanilla RS in terms of performance, particularly on the cold user subset. This indicates that Social4Rec is more effective in handling the challenges posed by cold users. 

From Table \ref{tab:offline}, we can draw the flowing findings:

\begin{itemize}
    \item We observe a substantial performance difference between the two datasets, highlighting the significant challenge faced by recommendation systems when dealing with cold users ($\textbf{RQ1}$). The performance margin indicates that cold users, who have limited or no interaction history, pose a greater challenge in accurately recommending relevant items.
    \item Social4Rec could significantly enhances the performance of existing recommendation system ($\textbf{RQ2}$), even with a simple aggregation method for social graph (i.e., Social4Rec$^{-}$).
    \item Comparing Social4Rec with Social4Rec$^{-}$, the experimental results in Table \ref{tab:offline} demonstrate the importance of the attention aggregator in Social4Rec for achieving overall better performance. The introduction of the social graph into the recommendation system is deemed appropriate. This suggests that social data, when guided by behavioral data, plays a crucial role in extracting interest characteristics and addressing the \textbf{Trivial Relation Challenge}.
    \item Among the variations of the Social4Rec model, Social4Rec$^{\dagger}$ experiences the most significant performance drop. This indicates that users with sparse relations cannot be properly assigned to the desired interest group, thereby significantly compromising the model's performance. The \textbf{Sparse Relation Challenge} is effectively addressed through the inclusion of the calibrator layer, which helps mitigate the impact of sparse relations on the recommendation quality.
\end{itemize}

Overall, the results emphasize the superiority of the Social4Rec model over Vanilla RS, especially in handling cold users. The attention aggregator, the utilization of social graph data, and the calibration of sparse relations contribute significantly to the improved performance.

To provide a deeper analysis of the performance of Social4Rec in the online system, we conducted experiments on a real-world video recommendation platform that boasts nearly one billion users worldwide. The experimental results were reported for two types of users: all users and cold users, using the same settings as the offline dataset.

Table~\ref{tab:online} presents the mean performance gains of Social4Rec compared to the best online model over a span of seven consecutive days, which provides more statistically significant results. From Table~\ref{tab:online}, we can obtain the following conclusions:

\begin{itemize}
    \item The overall Click-Through Rate (CTR) shows a remarkable improvement of 3.63\% when using Social4Rec. This improvement is highly significant in the context of industrial recommendation systems, indicating the efficacy of Social4Rec in enhancing user engagement and interactions.
    \item  Comparing the improvement in CTR for all users to that of cold users, we observe that Social4Rec achieves a performance gain of 2.00\% specifically on cold users. This further emphasizes the challenges faced by recommendation systems in effectively catering to cold users ($\textbf{RQ1}$). Cold users, who lack sufficient interaction history, present a more demanding scenario for recommendation algorithms, and the significant improvement on cold user segment highlights the potential of Social4Rec in addressing this challenge.
    \item Despite the challenges associated with cold users, the metrics of click number and view time exhibit significant improvements compared to the online model. This indicates that integrating the social graph into the recommendation system through Social4Rec enables a better personalized experience for cold users and encourages their conversion into active users to a certain degree ($\textbf{RQ3}$). 
    Active users exhibit a higher frequency of browsing video covers, clicking on and watching video content, which is observed in our video platform.
    This conversion of cold users into active users is vital for the continuous growth and development of the content platform.
\end{itemize}

Overall, the experiments conducted on the real-world video recommendation platform demonstrate the effectiveness of Social4Rec in improving the Click-Through Rate. 
By incorporating the social graph, the recommendation system provides a more personalized experience to cold users, facilitating their conversion into active users. These findings highlight the importance of Social4Rec in addressing challenges related to cold users and promoting the continuous growth and success of the content platform.

\begin{table}[!t]
\small
\centering
\caption{Offline performance comparison on two datasets. The relative performance improvement is statistically significant with $p <0.01$ in two-tailed paired t-test.}
\begin{tabular}{l|cc|cc}
\hline
  & \multicolumn{2}{c|} {\emph{FULL}} & \multicolumn{2}{c} {\emph{COLD}}  \\
  \cline{2-5}
  & AUC& Imp.\%& AUC& Imp.\%\\
\hline
Vanilla RS & 0.765&-&0.729&-\\
\hline
Social4Rec$^{\dagger}$ &0.767 &0.26\%& 0.735&0.82\% \\
\hline
Social4Rec$^{\ddagger}$ & 0.768 &0.39\%& 0.741&1.65\% \\
\hline
Social4Rec$^{-}$ & 0.768&0.39\%&0.739&1.37\% \\
\hline
\textbf{Social4Rec} & \textbf{0.770}&0.65\%&\textbf{0.746}&2.33\% \\
\hline
\end{tabular}
\label{tab:offline}
\end{table}


   
\begin{table}[!t]
\small
\centering
\caption{ Online performance gains over baseline model. The relative performance improvement is statistically significant with $p <0.01$ in two-tailed paired t-test.}
\begin{tabular}{lccc}
\hline
User type  & CTR  & click number  & view time   \\
\hline
All users    & +3.63\%  & +2.94\% & +0.78\%      \\
Cold users  & +2.00\%  & +8.59\%  & +4.77\%     \\
\hline
\end{tabular}
\label{tab:online}
\vspace{-10pt}
\end{table}

\section{Conclusion}
In this work, we introduce Social4Rec, a recommendation model framework that utilizes a social graph to mitigate the issues of sparse and noisy user behavior encountered by vanilla RS. The main objective is to design a framework that offers a more accurate and comprehensive representation of users, especially for new and cold users.

To address the challenge of sparse relations that arises from the introduction of the social graph, we propose a cluster-calibrate-merge module. The objective of this module is to identify intrinsic, robust, and self-consistent interest groups within the user population. Through clustering users based on their behavioral patterns and calibrating the sparse relations, we can effectively identify interest groups that capture relevant user preferences.

To tackle the challenge of trivial relations, we enhance the user behavior embedding by incorporating the user's interest group. This refinement process improves the quality and relevance of the user behavior representation. Furthermore, we employ an attention mechanism to aggregate relation-specific refined embeddings, resulting in an interest-aware social embedding. This embedding captures the user's social preferences, which are then combined with the user behavior embedding to compute the relevance score for each candidate item.

Our model achieves a more accurate and comprehensive representation of users by incorporating interest preferences derived from the social graph. This is particularly advantageous for new and cold users with limited behavioral data. Experimental results demonstrate the superiority of our Social4Rec model over the best online model in our video platform. The performance improvements in user engagement metrics underscore the effectiveness of incorporating the social graph in the recommendation process.


\balance
\bibliographystyle{splncs04}
\bibliography{splncs04}

\end{document}